# Interaction of Active Janus Particles with Passive Tracers

Karnika Singh[a], Ankit Yadav[a], Prateek Dwivedi[a] and Rahul Mangal[a,*]

[a]*Department of Chemical Engineering, Indian Institute of Technology Kanpur, Kanpur-208016, India*

*) *Author to whom correspondence should be addressed: mangalr@iitk.ac.in*

## Abstract

In this study, we investigated the motion of active $SiO_2$-*Pt* Janus particles in the 2D bath of smaller silica tracers dispersed in varying areal densities. The effect on the organization of the tracer particles around the active JPs was also explored. Our experiments indicate that the interaction between the tracers and the active JPs mainly depend on the nature of collision marked by the duration of contact. For all the concentration regimes, we have shown that the short time collisions do not have significant impact on the motion of active JPs, however, during moderate/long-time collisions tracer(s) can lead to a significant change in active JPs' motion and even cause them to rotate. In the concentrated regime, our experiments reveal the emergence of a novel organizational behavior of the passive tracers on the trailing *Pt* and the leading $SiO_2$ with a strong dependence on the nature of collision.

## 2 Introduction

Artificial active matter systems demonstrate the capability to perform non-equilibrium motion by utilizing the energy from their surroundings (*1, 2*). In doing so, they successfully mimic the motion of several micro-organisms such as bacteria (*E. coli*), sperm cells, etc (*3*). Therefore, studying the isolated and collective motion of such artificial active systems provides useful insights into the motion of their biological counterparts. In addition, these artificial systems have shown the capability to be useful in several potential applications such as drug-delivery, micro-machines, environmental remediation, biological sensing and imaging etc (*4–8*). Artificial active matter systems can be broadly classified in two categories i.e. Active Colloids (ACs) and Active Droplets (ADs), based on their propulsion mechanism. For ACs among the various strategies that have been devised over the years, the use of Janus particles (JPs) has been widely recognized (*9*). The two opposite sides of the JPs (spheres or rods) are of different chemical compositions (*10*, *11*). To make them ACs one side is usually coated with a reactive metal such as *Pt* or *Pd*, and the other side is of a dielectric material such as $SiO_2$ or $TiO_2$ (*12*). The most commonly used candidates are $SiO_2$-*Pt* JPs. On subjecting these $SiO_2$-*Pt* JPs into an $H_2O_2$ aqueous solution, the *Pt* site promotes the decomposition of $H_2O_2$. The consequent anisotropic distribution of the reaction products generates a chemical potential gradient across the particle which generates motion in the surrounding fluid towards the *Pt* site (*13*, *14*) . In absence of any external force, to balance the overall linear momentum the JP gets propelled in the opposite direction, away from the *Pt* side. This propulsion mechanism, based on the chemical potential gradient, is known as diffusiophoresis and since the chemical gradient is generated by the particle itself, it is known as self-diffusiophoresis (*15–17*). In the case of active droplets, self-propulsion is a resultant of Marangoni flow in the surrounding fluid which originates due to gradient in concentration of surfactant molecules along the interface felicitated by the continuous dissolution of droplet molecules in the bulk liquid (*18*, *19*).

For their successful use in most of the sought-after applications, it is critical to gain an in-depth understanding of their motion, both isolated and collective, under varied scenarios. This includes presence of nearby obstacles, external-flow and inert impurities. In the bulk of a simple Newtonian liquid, several experimental and analytical studies has established that such active JPs perform ballistic motion at short time scales which transitions to a random motion at long time scales (*20*). This transition is prompted by the Brownian fluctuation in the orientation of the JPs. In the vicinity of an impenetrable wall, the complex interplay of the wall-induced hydrodynamic interactions, the chemical activity, and gravity, cause JPs to align with their Janus plane being perpendicular to the wall (*21–24*). Close to the wall, on exposure to shear flow, experiments have shown that these JPs can migrate upstream or perpendicular to the shear flow (*25–27*). The understandings of these wall-induced interactions have been further implemented to guide the motion of active JPs (*28*), capture the active JPs by stationary obstacles (*29*) and to sort them through well designed maze (*30*). So far very few studies have investigated the interactions of fast moving active JPs with mobile inert passive tracers (*31*). Most of these studies have discussed the 2D clustering behavior of the tracer particles with the light activated JP as the nucleating site (*32–34*), originated by the activity induced flow-fields in the vicinity of the wall. Recently, Hailing *et al* demonstrated the use of active JPs for directional locking and channel unclogging of silica beads in micro-channels (*35*). In a very recent study Katuri *et al* (*36*) by exploring the motion/structure of the passive tracers surrounding the active JPs, mostly immobile, map the activity induced hydrodynamic fields. The study indicated that the interactions between the tracers and active JPs are complex and dependent on the state of active JP. Given the underlying rich physics, further experimental studies are warranted to explore the complex interactions between the tracers and motile active JPs under different scenarios. In this study, we dispersed active $SiO_2$-*Pt* JPs in a bath of passive silica tracers under dilute, semi-dilute and concentrated regimes. The effect of passive tracers on the motion of active JPs and the complimentary effect on passive tracers' distribution is experimentally investigated.

# 3 Experimental Section

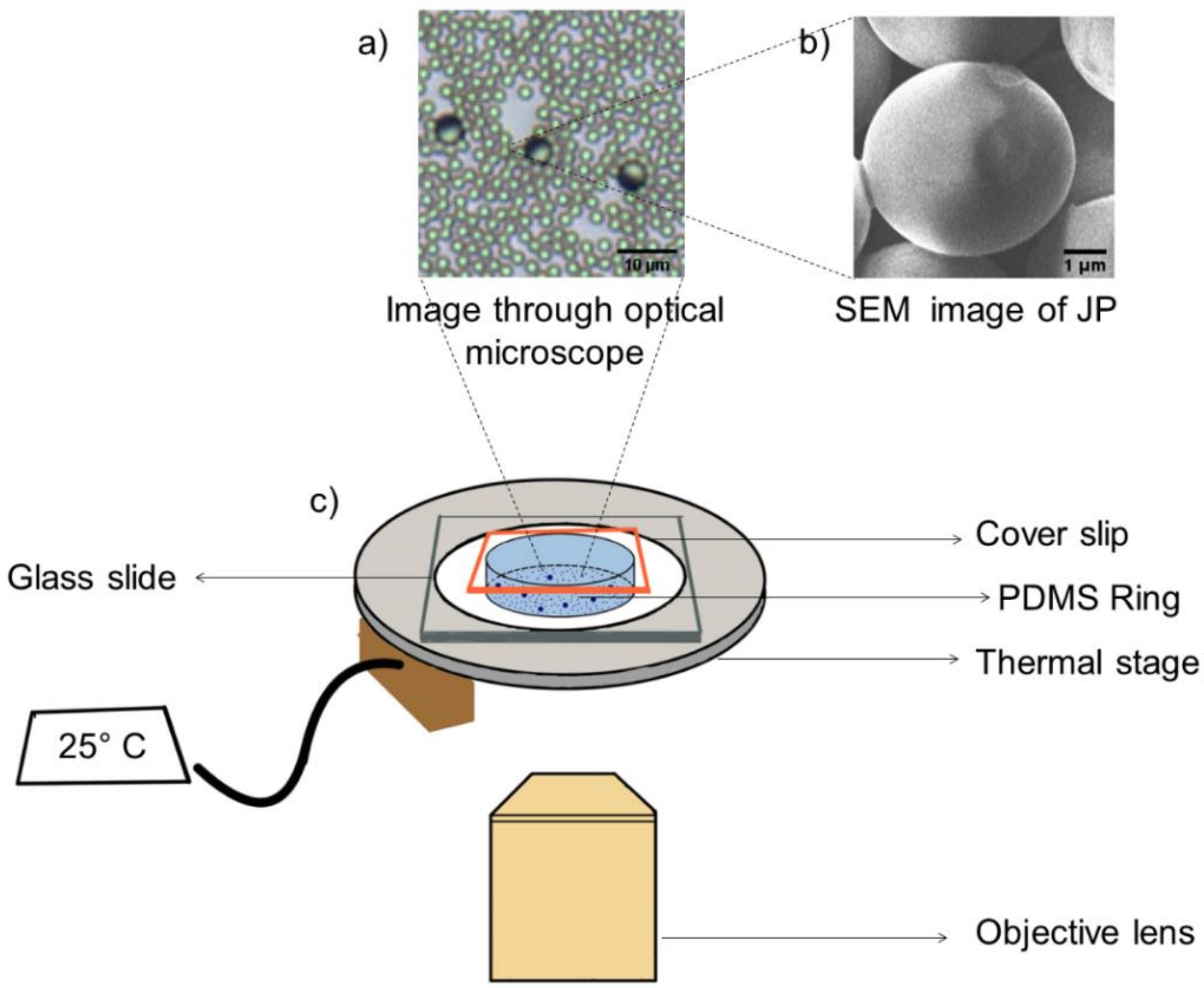

Figure 1: (a) Optical micrograph of the active JP in the passive tracer bath. Scale bar corresponds to 10 µm. (b) SEM micrograph of the $SiO_2$-Pt Janus colloid. Scale bar corresponds to 1 µm. (c) Schematic of the experimental set-up.

Janus Particles were synthesized using the drop-casting method as reported by Love *et al* (*37*). Briefly, a dilute suspension of 5 µm $SiO_2$ colloidal particles was prepared in Milli-pore water with a (1:3) ratio. The suspension was drop cast on a plasma-treated glass slide used as a substrate. Plasma treatment was done to make the glass slide hydrophilic which helps in the uniform spreading of the particle suspension. The water was removed via slow evaporation which ensured the formation of the monolayer formed, confirmed via optical microscopy as shown in figure 1. A thin layer of Platinum (~ 15 nm) was further deposited on the particles via plasma sputtering in a Sputter Coater. Due to the self-shadowing affects the top half of the particles only gets coated which make them Janus. Using sonication, thus prepared JPs were dispersed in DI water. The Optical micrograph shown in figure 1(a) and SEM micrograph shown in figure 1(b) illustrate the successful *Pt* coating and formation of Janus particles. To observe the particles, an optical cell was prepared by sandwiching two plasma-treated glass slides using a spacer (PDMS ring of height 2 µm). An appropriate amount of the 2 µm $SiO_2$ particle solution, used as passive tracers, was added to achieve different area fractions $\varphi_A = {Na_p}/{A}$, where $N$ is the total number of particles in the frame, $a_p$ is the area of an individual particle and $A$ is the total area. A small amount of the JPs solution was added to keep the JP concentration low. An appropriate amount of $H_2O_2$ aqueous solution was also added to achieve the overall $H_2O_2$ wt% as 0% and 3%. Finally, the optical cell was covered with a coverslip of (~ 100 µm). The optical cell was then placed on the thermal stage mounted on the microscope stage for -imaging. A constant temperature of 25°C was maintained using the thermal stage.

## 4 Methodology

Since the particles are heavy, they sediment towards the bottom wall of the optical cell and move in the 2D X − Y plane with insignificant movement in the Z direction. Particles were imaged using FLIR camera with resolution (1pixel= 0.22 µm) equipped with Olympus Inverted Microscope. To allow the system to attain a constant temperature of 25°C a waiting period of 10 minutes was used after placing the optical cell on the thermal stage. Subsequently, movies for 5 − 10 mins were recorded at 20 fps. Afterward, particle tracking was done using MOSAIC plugin in Image-J that uses a correlation-based approach to obtain particle trajectories. The trajectories were analyzed for 100 sec to get the time-dependent X-Y data.

For Brownian particles, the mean square displacement (MSD) $\langle \Delta L^2 \rangle$ is increasing linearly with time. The data were fitted using the equation $\langle \Delta L^2 \rangle = 4D\Delta t$. Here, D is the short time scale Brownian Diffusivity of the particle at temperature T in a liquid with viscosity $\eta$. For active JPs, the MSD is parabolic in nature which was fitted with using the equation

$$\langle \Delta L^2 \rangle = v^2 \Delta t^2 + 4D\Delta t \ldots\ldots\ldots\ldots(1)$$

where v is the average propulsion speed v of the active JPs. The instantaneous velocity $\mathbf{v}_{inst.} = \dfrac{\mathbf{r}_{i+1} - \mathbf{r}_i}{t_{i+1} - t_i}$ of the active JPs was used to compute the velocity autocorrelation function $C(\Delta t) = \langle \mathbf{v}_{inst.}(\Delta t) \bullet \mathbf{v}_{inst.}(0) \rangle$ which was fitted using the equation:

$$C(\Delta t) = 4D\delta(\Delta t) + v^2 \cos(\omega \Delta t) \exp\left(\frac{-\Delta t}{\tau_R}\right) \ldots\ldots(2)$$

Here $\tau_R$ is the rotational time scale, δ is the Dirac-delta function, and ω is the angular velocity of the active JPs to capture the curvature in the trajectories, if any. A representative plot for $C(\Delta t)$ with fitting has been shown in supporting figure 2.

The distribution of passive particles around the JPs was computed using the 2D radial distribution function $g(r) = \dfrac{\rho_{local}}{\rho_{bulk}} = \dfrac{\left( n / 2\pi r dr \right)_r}{\left( N_{total} / A_{total} \right)}$, here $n$ is the passive particle count at distance $r$ away from the surface of the JP. $N_{total}$ is the overall number of passive particles around the JP in a total area of $A_{total}$.

## Results and Discussion

### Control Experiments

To first establish the validity of our experimental setup and protocols, we performed some control experiments, as discussed below:

#### Passive Motion

Appropriate amounts of 5 μm $SiO_2$-*Pt* JP suspension and 2 μm $SiO_2$ tracer particle suspension were mixed to achieve $\varphi_A$ ~ 0.35 for the 2 μm $SiO_2$ particles and $\varphi_A$ ~ 0.06 for the 5 μm $SiO_2$-*Pt* JPs. In absence of $H_2O_2$, the JPs perform 2D Brownian motion which was analysed using their 2D

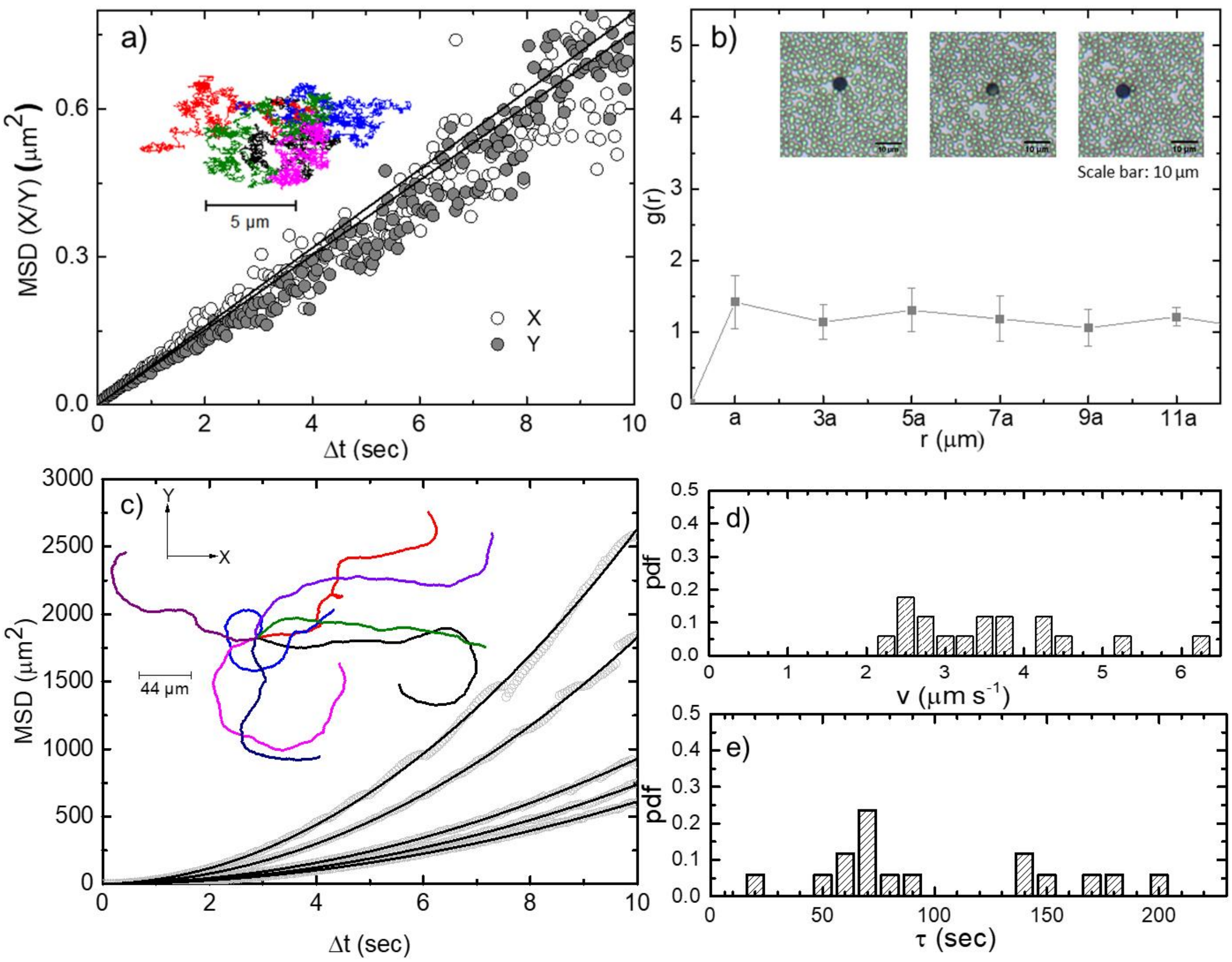

Figure 2: (a) MSD (X,Y) of a passive JP. Solid lines are the linear fit. Inset shows few representative X-Y trajectories. (b) Radial distribution function for passive tracers around few JPs. Inset shows the representative optical micrographs. (c) Representative MSDs (X,Y) of active JPs in absence of passive tracers. Solid lines are the fit to equation 1. Inset shows few representative X-Y trajectories. (d),(e) probability distribution of propulsion speed and rotational time scale for the system.

X −Y trajectories and the corresponding mean square displacement (MSD) data as shown in Fig 2(a).

As expected, the trajectories are near isotropic in nature without any preference to X or Y direction and the corresponding mean square displacement MSD(X)~MSD(Y), confirming negligible convection effects in the system. Also, consistent with Brownian motion the MSD(X,Y) increases linearly with time. The slope of the curves which is the short time tangential diffusivity $D$ of the Janus particle $D(\mathrm{X}) \sim D(\mathrm{Y}) \sim 0.0389 \pm 0.01\ \mu\mathrm{m}^2\,\mathrm{s}^{-1}$ which is lesser than the expected Stokes-Einstein prediction $D_{SE} = (kT/6\pi\eta a) = 0.09\ \mu\mathrm{m}^2\mathrm{s}^{-1}$. This reduced tangential diffusivity of the passive JPs is due to their interaction with the surrounding passive tracers and the bottom wall. It was also seen that in absence of any activity, due to the density difference in the *Pt* and the $SiO_2$ side, the Janus particles orient with the *Pt* side facing downwards. Next, to understand the distribution of tracer particles around the JP, we compute the 2D radial distribution function g(*r*). Figure 2(b) indicates that close to the JP at $r$ =a; g(r) peaks (> 1) which further drops down and saturates at 1, away from the JP.

**Active Motion in absence of passive tracers**

Next, the motion of isolated 5 µm active JPs in absence of passive tracers was investigated in presence of 3 wt% $H_2O_2$. The propulsion arises from the decomposition of $H_2O_2$ on the Pt side generating a chemical gradient across the particle leading to self-diffusiophoretic motion of the particle. Few representative 2D X−Y trajectories and the corresponding mean square displacement (MSD) data have been shown in figure 2(c). As expected, the trajectories are near isotropic in nature without any preference to X or Y direction. Also, unlike Brownian motion, the MSD plots of these active JPs are parabolic, which is indicative of their self-propelled nature. By documenting the behaviour of 20-30 active JPs, in figures 2(d) and 2(e) we show the distribution of the computed values of the propulsion speed v and rotational time scale $\tau_R$. The distribution in the values underscores the inherent heterogeneity in the motion of the active JPs which is generally attributed to the non-uniformity in Pt coating, size distribution, and wall interactions. Another key observation from the recorded videos is that mostly the orientation of the JP is such that their normal to the Janus plane (**n**) is parallel to the bottom wall. This stable orientation of active JPs has been attributed to the interplay of the hydrodynamic and the chemophoretic interactions with the wall. (*28*, *38*)

**Active JPs in a dilute suspension of passive tracers**

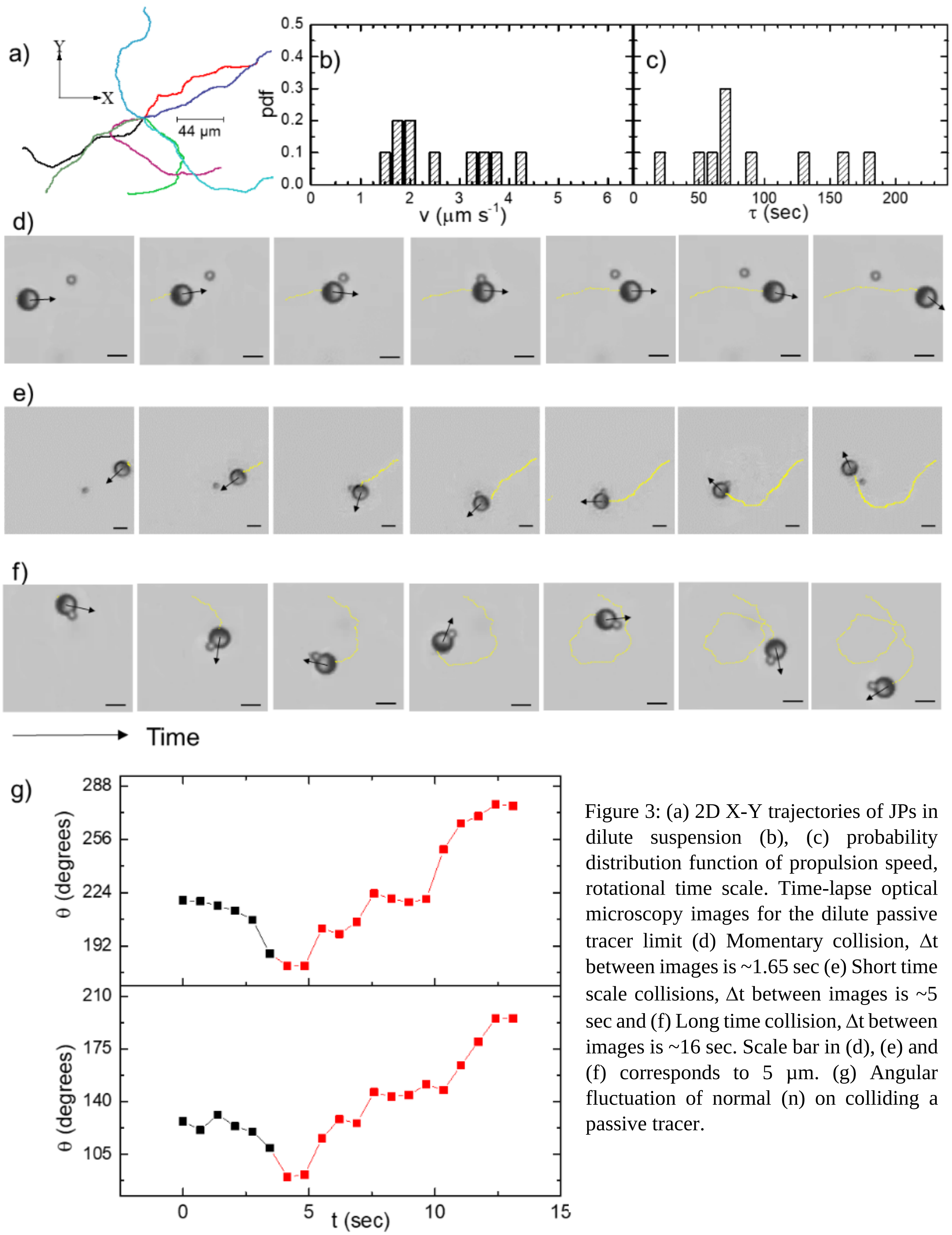

Figure 3: (a) 2D X-Y trajectories of JPs in dilute suspension (b), (c) probability distribution function of propulsion speed, rotational time scale. Time-lapse optical microscopy images for the dilute passive tracer limit (d) Momentary collision, Δt between images is ~1.65 sec (e) Short time scale collisions, Δt between images is ~5 sec and (f) Long time collision, Δt between images is ~16 sec. Scale bar in (d), (e) and (f) corresponds to 5 µm. (g) Angular fluctuation of normal (n) on colliding a passive tracer.

Next, we explore the effect of passive tracers under a dilute regime ( $\varphi_A$ ~ 0.03) on the motion of active JPs. Figure 3(a) illustrates few representative 2D X-Y trajectories. In Figures 3(b) and 3(c), we show the distribution of the propulsion speed computed from the X-Y data. The data clearly suggests that under a dilute regime the passive tracers only have a slight effect in reducing the propulsion sped v and the $\tau_R$. Due to the lower density of passive tracers the active JPs propel freely and remain mostly undisturbed and very occasionally face a collision with silica tracers. Careful inspection of the recorded videos reveals that such occasional collisions of the active JPs with the passive tracers varied in their time for which the particles remain in physical contact and thus can be broadly classified in three categories: (i) short time-scale collision (ii) moderate time-scale collision and (iii) long time-scale collision. During short-time scale collisions, the passive tracer and the active JP maintain physical contact for a very short time scale ~ 1.65s and immediately after the collision the particles move away. During such a collision no significant change in the motion (speed and direction) of the active JP is noticed. A typical sequence of images for such a collision is shown in figure 3(d) (see also Video S2 in the Supporting Information). During moderate time-scale collisions, physical contact is maintained for slightly longer durations ~ 5 sec. On being approached by an active JP, the flow-field generated by the active JP advects/pulls the tracer particle towards the $SiO_2$ side. Once a close contact between the two is established, due to the strong hydrodynamic interactions, instead of bouncing away from the JP, the tracer continues to advect with the flow-field while maintaining contact with it. Subsequently, it slides along the JPs' surface towards the equator side and detaches upon entering the *Pt* region. Figure 3(g), illustrates the time evolution of the direction of **n** of active JPs during the process of two such collisions. Clearly, just before the collision the JP rotates away from the tracer particle, after which they rotate towards the tracer particle until the tracer detaches. This behavior is also evidenced in representative time-lapse images shown in figure 3(e) and video S3 in the Supporting Information). During the initial approach, even when the particles are far-apart from a physical contact, the long ranged hydrodynamic interactions emerge. The tracer particle successfully modifies the flow-field in front of the active JP, causing it to rotate away from the passive tracer. This observation agrees with the recent analytical predictions made by Purushothaman and Thampi (*39*). Subsequently, when the passive tracer is attached to the active JP, it generates an asymmetry in the otherwise symmetric flow-field around the JP, causing the particle to rotate in the direction towards the tracer particle to conserve the overall angular momentum.

Our results indicate that despite the significant size difference and overall low concentration, smaller $SiO_2$ particles no longer just act as tracer particles, instead, remarkably affect the motion of an active JP. The rotational behavior is similar to the previously reported study on smaller active JPs near the fixed obstacles. However, for the first time to the best of our knowledge, we have demonstrated the influence of a smaller movable tracer particle significantly affecting the motion of a bigger active JP. Finally, during the long time-scale collisions, the passive tracer remains stuck with the active JP after colliding, for very long times. Such active JPs continue to perform the circular motion, where the direction of rotation is again towards the side where the tracer is attached (c.f. representative time-lapse images are shown in figure 3(f)). The representative movie has been shown in supporting video S4.

**Active JPs in the semi-dilute suspension of passive tracers**

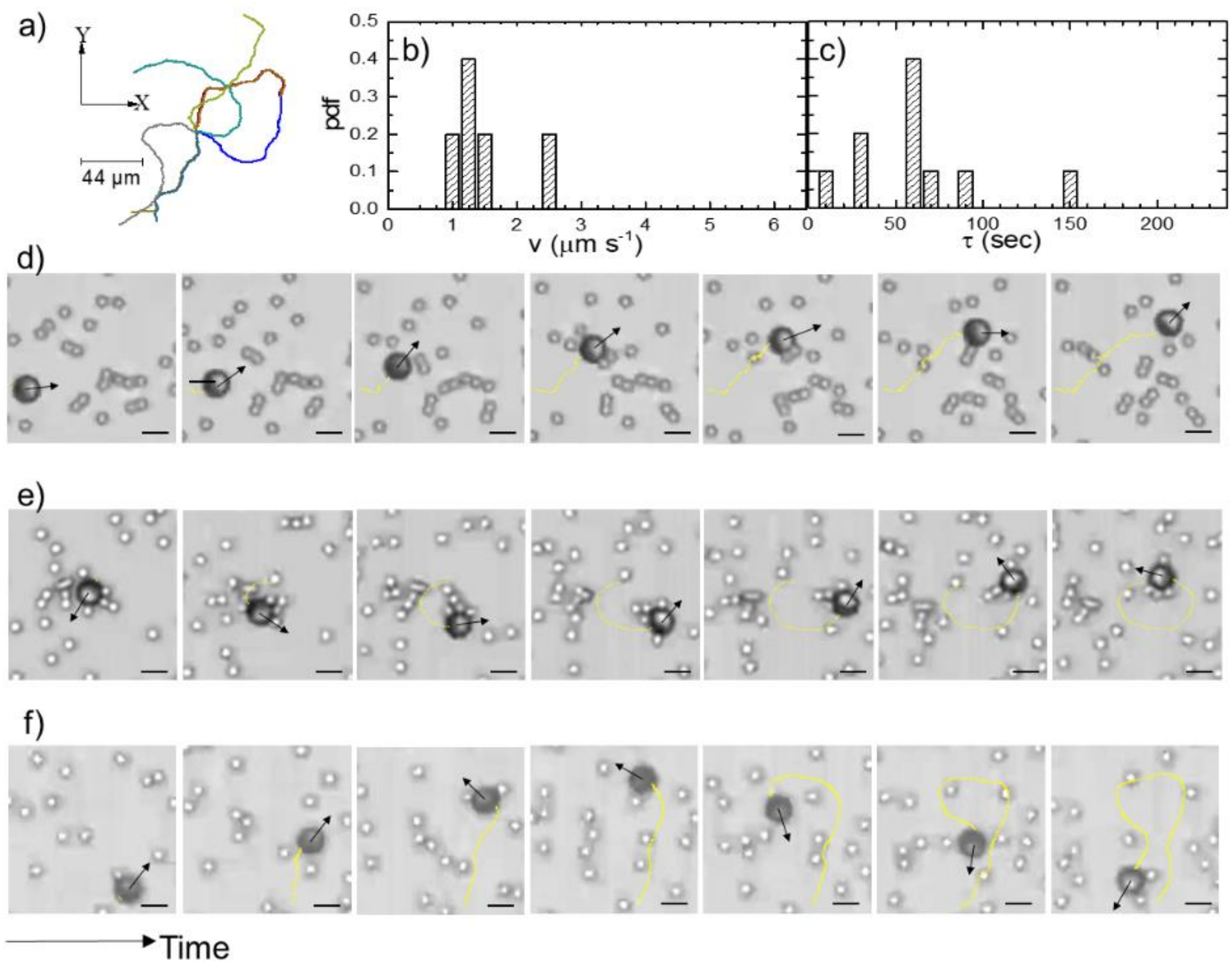

Figure 4: (a) 2D X-Y trajectories of JPs in semi-dilute suspension (b), (c) probability distribution function of propulsion speed, rotational time scale. Time-lapse optical microscopy images for the dilute passive tracer limit (d) Moderate time collision Δt between images is ~ 9 sec (e) Long time collision, Δt between images is ~ 21 sec (f) Momentary collision, Δt between images is ~ 16 sec. Scale bar in (d) and (e) corresponds to 5 μm.

Subsequently, we explored the effect of passive tracers under a semi-dilute regime ($\varphi_A$ ~ 0.19) on the motion of active JPs. Figure 4(a) shows few representative 2D X-Y trajectories. In figure 4(b) and 4(c) we show the variation in the computed speed and rotational time scale for an ensemble of 20-30 active JPs. The data clearly suggests that under the semi-dilute regime, due to the enhanced probability of the collisions with the passive tracers, both the speed and $\tau_R$ of the active JPs are noticeably less compared to the tracer-free case. Similar to the dilute regime, the nature of collisions remains the same. However, unlike the previous case, in this regime multiple tracers collide with multiple JPs at the same time, therefore the effect on the motion is more pronounced. (c.f. representative time-lapse images are shown in figure 4(e)). Representative movies have been shown in supporting movies S5-S6.

**Active JPs in concentrated suspension of passive tracers**

Finally, the motion of active JPs was also investigated in a dense bath ($\varphi_A$~0.35) of passive tracer particles. Figure 5(a) shows few representative trajectories captured for different active *JPs*.

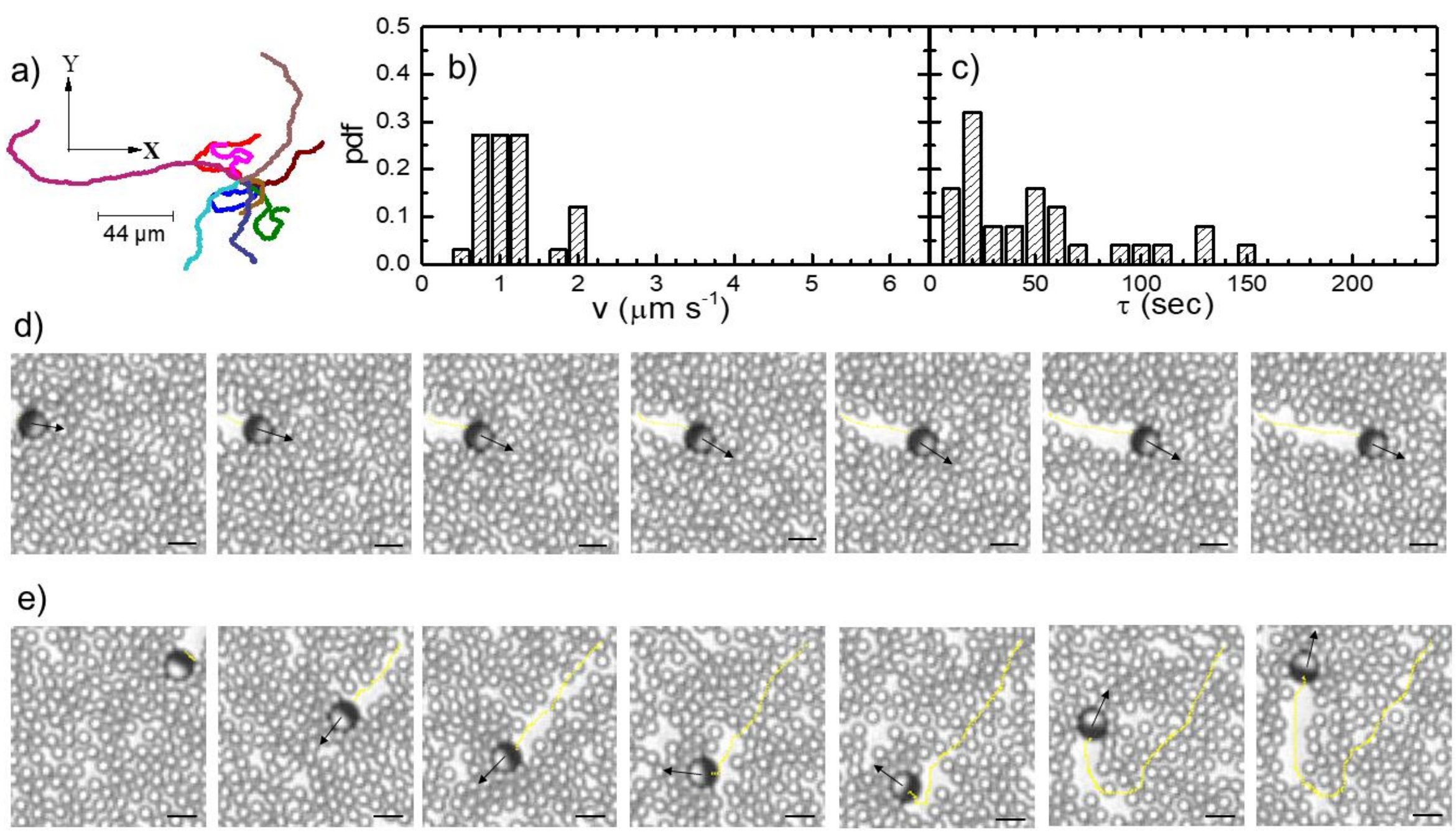

Figure 5: (a) 2D X-Y trajectories of JPs in concentrated suspension (b), (c) probability distribution of propulsion speed and rotational time scale (d) Moderate time collision, Δt between images is ~ 2 sec (e) Long time collision, Δt between images is ~ 13 sec. Scale bar in (d) and (e) corresponds to 5 µm.

By documenting an ensemble of 20-30 active JPs in figure 5(b) we show the distribution of their speed. The data clearly suggests that in comparison to the control case, the presence of numerous passive tracers significantly slows down the overall motion of the active JPs. Figure 5(c) illustrates the distribution of the rotational time-scale $\tau_R$, wherein, a shift in $\tau_R$ towards lower values, in comparison to the control case is observed. These results indicate that due to the repeated encounters of the active JP with the passive tracers in the dense bath, the active JPs not only feel enhanced hindrance to move but are also forced to change their direction more frequently. In such a crowded environment, collisions with passive tracers are mostly either moderate-time or long-time collisions. During moderate-time collisions, the passive tracers while maintaining contact with the JP slide along the JPs' surface towards the Pt side, where they detach. Therefore, the immediate layers of passive tracers around the JP keeps changing. However, unlike previous regimes where the collisions lead to asymmetric phoretic slip around the JP, in a concentrated system, for moderate-time collisions, the asymmetry is significantly minimized (c.f. see representative time-lapse images shown in figure 5(d)). Therefore, the deflection in the motion of the active JP is not noticeable. However, in some cases, few tracer particles remain stuck to the active JP, especially the closest few layers close to the JP, causing the particle to significantly rotate and change directions, see representative time-lapse images shown in figure 5(e). Representative movies have been shown in supporting movies S7-S8.

Further, from the supporting movies S7-S8 and the time-lapse snapshots shown in figure 5(d) and 5(e), it can be seen that when the active JPs move through the pool of passive tracers, they develop a wake i.e. a depletion zone of passive tracers at their trailing *Pt* side. In a very recent publication, Katuri *et. al* demonstrated a similar wake near the *Pt* side (*36*) which has no specific shape. However, in their experiments, the active JPs were immobilized and hence the wake was attributed to the decomposition of $H_2O_2$ on the *Pt* side. In contrast to their observations, in our experiments, the active JPs are mobile which results in few major differences. In figure 6(a), we demonstrate the variation in the average wake area $\langle A_{wake} \rangle$ with an average instantaneous speed $\langle |v_{inst.}| \rangle$ of the active JPs. Here, $\langle \ \rangle$ represents the time and ensemble average. The speed variations in the active particles are inherent to the system and are present even in the control case also, as discussed earlier. The figure suggests that despite a fixed $H_2O_2$ concentration, $\langle A_{wake} \rangle$ increases nearly linearly with $\langle |v_{inst.}| \rangle$. However, when the active JPs are at a halt i.e. $|v_{inst.}|$~0, $A_{wake}$ ~20 $\mu m^2$ which is approximately similar to the area reported by Katuri *et al.*(*36*) for immobilized JPs. However, when

$|v_{inst.}|$ >>0, despite the same 3wt % $H_2O_2$ concentration, the wake area (~ 90 μm$^2$) in our experiments was found to be much more than 20 μm$^2$ reported by Katuri et al.

We also observed that during the motile state of the active JP, the wake resembles appearance with the comet-like pattern, wherein it is elongated along the velocity direction of the particle. On the other hand, during the rest state the wake resembled a fluid-like shape reported by Katuri et al. These differences emphasize that besides the chemical activity of the active JPs, their translation motion also plays a crucial role in enlarging the wake area. To further understand this, we investigated the wake disappearance time under different conditions. The plot shown in figure 6(b), indicates that for moderate time collisions where the tracer particles glide over the active JP and detach, the wake disappears in nearly similar time ~10s irrespective of the area of the wake. Figure

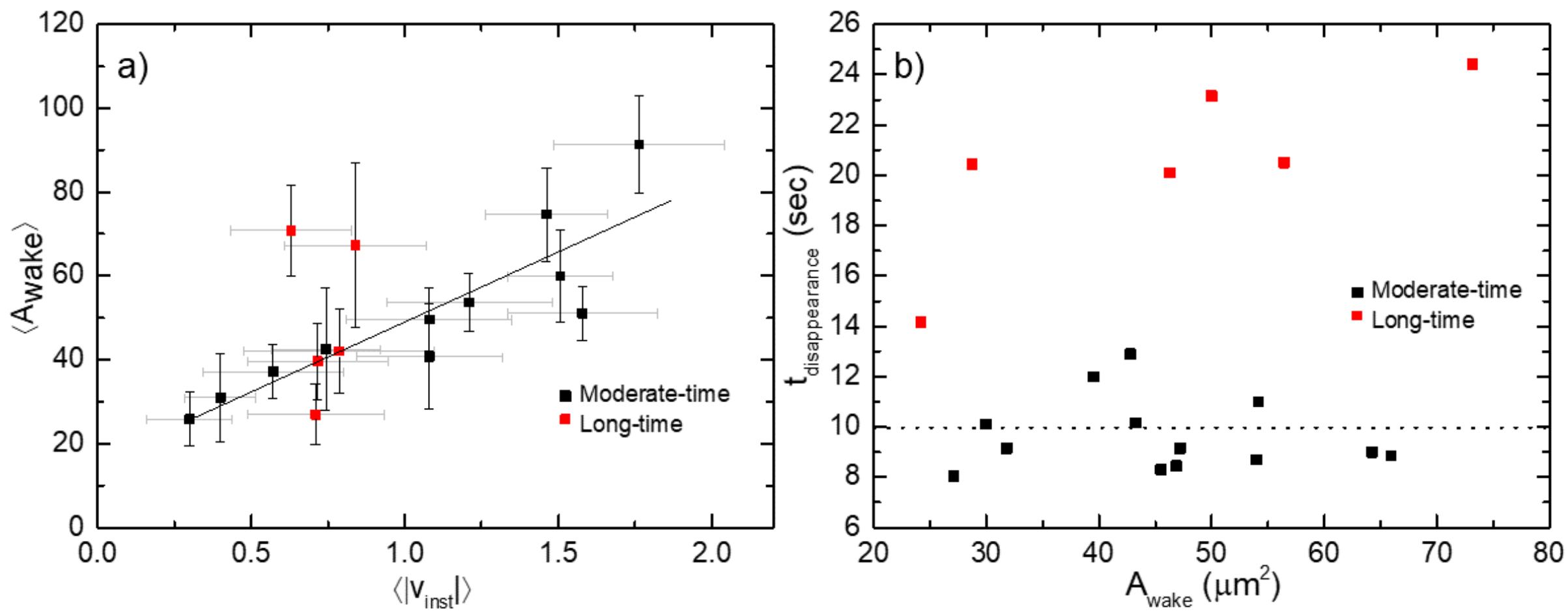

Figure 6: (a) Average wake area with speed for moderate and long-time collision. The solid line is mere guide to the eye. (b) Disappearance time with wake area

7(a) and 7(b), shows the time-lapse images of a wake disappearing gradually. The sequence of images indicates that as the active JP moves out of the wake, the peripheral tracer particles along the longer edges move perpendicular to the length of the wake to fill the void.

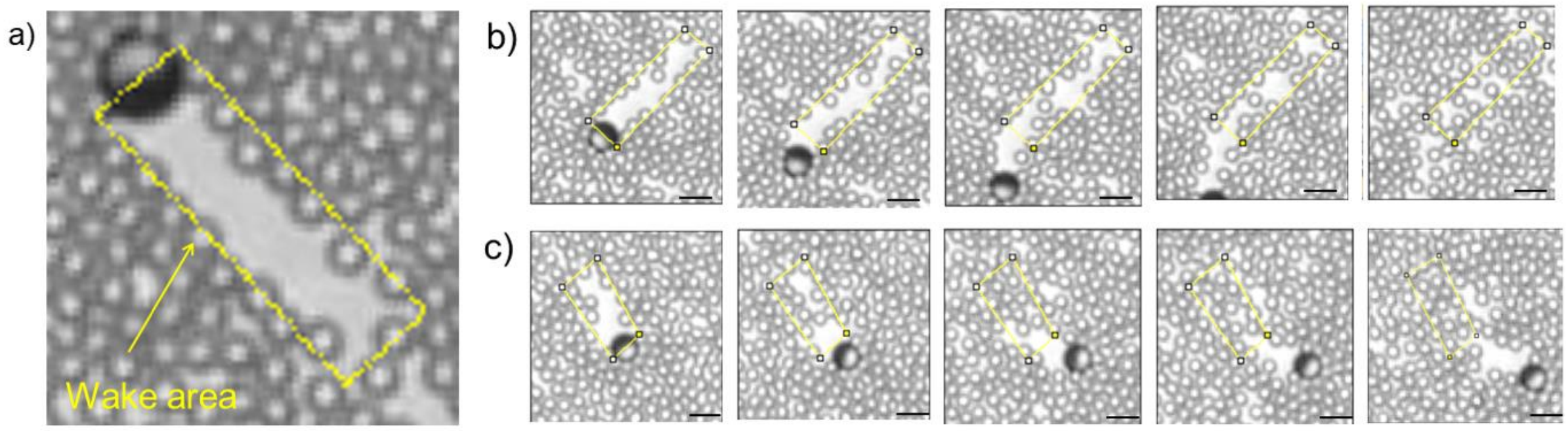

Figure 7: (a) Representation of wake area. Time-lapse optical microscopy images for the wake area disappearing in concentrated passive tracer limit (b) Moderate time collision, Δt between images is ~2.21 sec, wake disappearance time ~8.85 sec (c) Long time scale collisions, Δt between images is ~5 sec, wake disappearance time ~20.1 sec. Scale in (b) and (c) corresponds to 5 μm.

To confirm this, we performed a simple back of the envelope calculation where we assume the wake to be as a simple rectangular area with the width W (~5 μm) and of a certain length which depends on the active JP speed. The diffusive time scale pertinent for wake disappearance can be estimated as $t \sim \frac{(W/2)^2}{D_{SE,tracer}} \sim 10s$, where W/2 is the lateral distance to be covered by tracer on each side and $D_{SE,\ tracer} \sim 0.25\ \mu m^2\ s^{-1}$. The close match of the experimental time with our computation confirms that due to the motility of the active JPs the depletion region is generated which the slow diffusing passive tracers are unable to fill instantly hence a void is developed. On the other hand, for the wakes formed by the long-time collisions, where the tracer particles remain stuck for longer times, the wakes disappear slowly (figure 7(b)) as the JP moves out of the area slowly due to its lesser speed. In the limiting case where the active JP are stuck, we expect the wake to persist as long the particle remains active.

The emergence of a wake on the *Pt* side is accompanied by the accumulation of passive tracers at the leading edge of the active JPs. We quantity this by computing the variation in g(r) at the front half. In Figures 8(a) and 8(b), we demonstrate the average g(r) for the passive traces on the silica side about the center of an active JP. Figure 8(a) indicates that in the case of moderate-time collisions, the first peak at r=*a* of g(r) is significantly higher than the control case. The increase in the second peak at r=3*a* is not much and eventually at r=7*a* the g(r) nearly matches with the control case (here a is radius of the passive particle and r is measured from the surface of the JP). In case of long-time collisions, it is observed that similar to the moderate-time collision although at r=7*a*, g(r) approached the control case values, however the height of the second peak at r=3*a* is more pronounced. This confirms that the nature of collision affects the accumulation of passive tracers at their leading also.

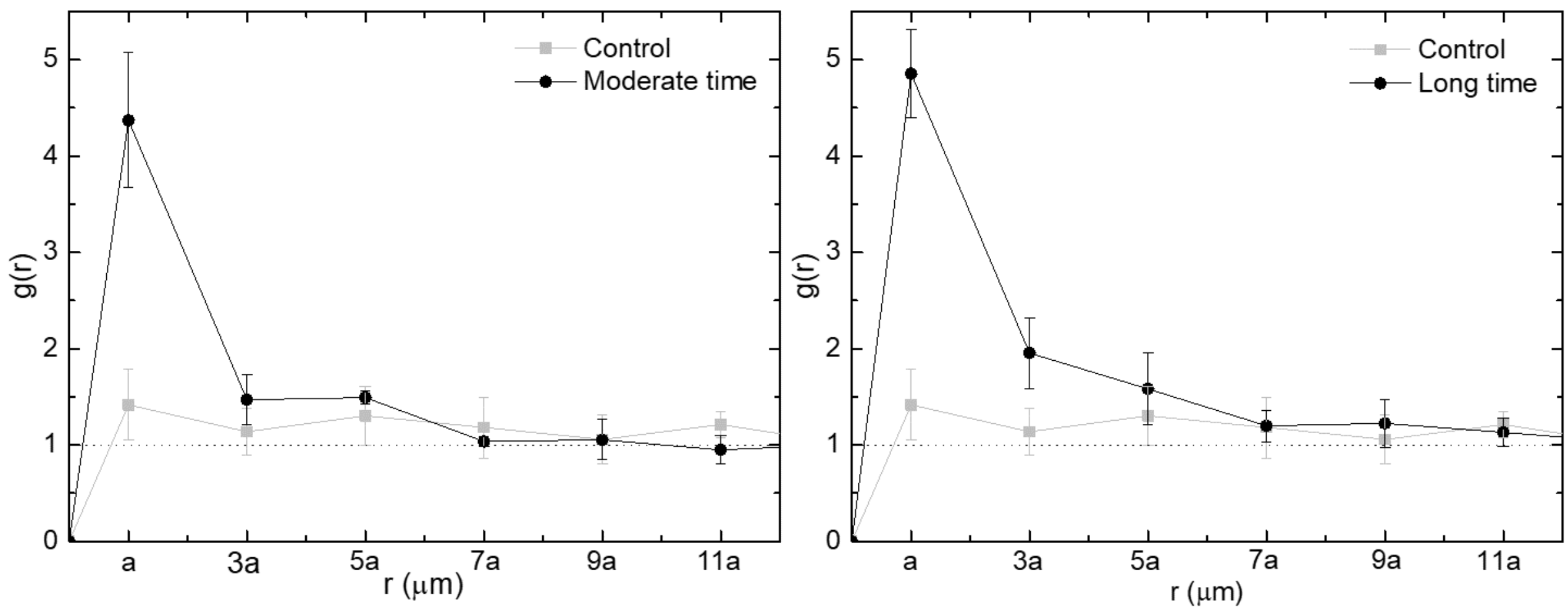

Figure 8: Radial Distribution function (a) Moderate time collisions (b) Long time collisions

Although the areal coverage of passive traces in the concentrated regime is not very high to achieve the jammed state. However, occasionally it was observed that when two active JPs moved in a manner which resulted in a finite approach velocity along the line joining their centers, the passive tracers in the intermediate region approached the jammed state. This local jamming of the tracer particles induces a long-range interaction between the active JPs forcing them to turn and change their direction of motion. As a result, the two JPs are unable to come in physical contact with each other, which however in the control case, in absence of intermediate passive tracers the active JPs are able to touch most of the time. This observation is further highlighted in the minimum approach distance $R_{\min}$ between the active JPs in the two scenarios, shown in figures 9(a) and 9(b). In the concentrated regime, $R_{\min}$ is much larger than the $R_{\min}$ in the control case.

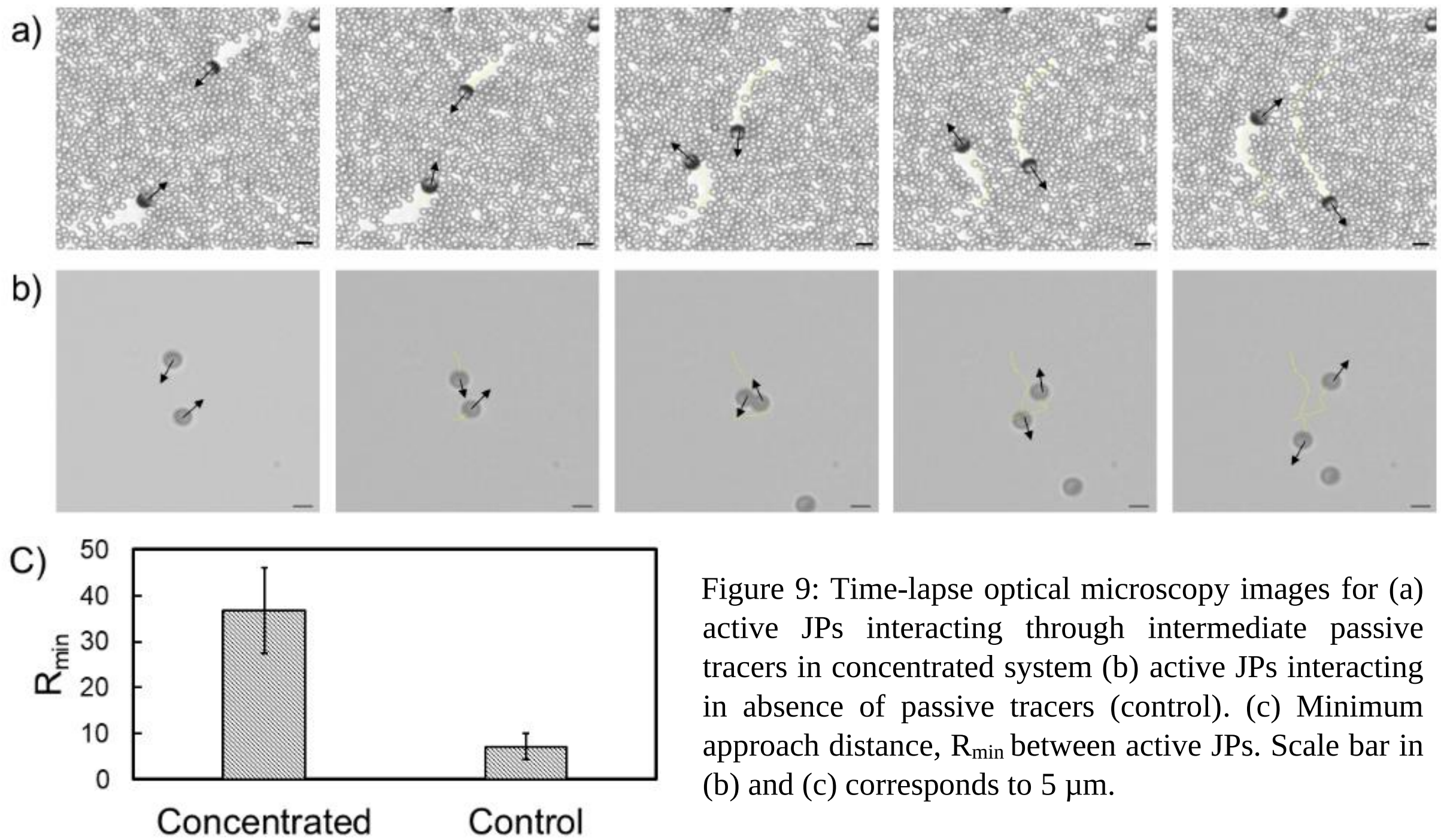

Figure 9: Time-lapse optical microscopy images for (a) active JPs interacting through intermediate passive tracers in concentrated system (b) active JPs interacting in absence of passive tracers (control). (c) Minimum approach distance, $R_{min}$ between active JPs. Scale bar in (b) and (c) corresponds to 5 µm.

## Conclusions

In this paper, we have experimentally discussed the effect of a 2D bath, with varying areal density, of passive $SiO_2$ tracers on the motion of active (using $H_2O_2$) $SiO_2$-Pt JPs and the organization of the surrounding tracers. Our observations conclude that the collisions of the active JP with the passive tracers, which are marked by their time for which the particles maintain the physical contact, have significant impact on the dynamics (speed and rotational time) of the swimmer. In the dilute tracer regime, presence of lower number of passive tracers, results in less collision frequency with active JP which doesn't alter their speed noticeably. However, the collision alters the hydrodynamic flow-field around the JPs, causing them to rotate. In semi-dilute regime higher number of collisions with passive tracers resulted in lower speed and more rotations. In concentrated bath of passive tracers, higher frequency of collision results in significantly suppressed mobility of the active particles with lower rotational time. The motion of active JP in the concentrated pool of passive tracers results in a denser region of pool of passive tracers in the front $SiO_2$ side and the depleted region at rear Pt side of active particle. Another key observation was that depending on the type of collision the depleted region or the wake region demonstrated varied dependence on the average instantaneous speed of the active JP.

We also concluded that the lateral Brownian diffusivity of the passive tracers primarily lead to the disappearance of the wake, which was further supported by the observation that the wake disappearance time was found to be independent of the wake area. Additional experiments and deeper analytical studies will bring more understanding on these passive crowded systems.

## Supplementary Materials

MSD for the JPs in the dilute, semi-dilute, and concentrated regime, Movies for the experiments: S1- Motion of JP in 3% $H_2O_2$(Control), S2- S4- Short time, moderate and long-time collision of JP in a dilute suspension of passive tracers, S5-S6- Moderate and long-time collision of JP in the semi-dilute suspension of passive tracers, S7-S8- Moderate and long-time collision of JP in a concentrated suspension of passive tracers, S9- Closest separation distance between JP in control S10- Closest separation distance between JP in a concentrated suspension

### Author's Contributions

RM conceptualized the problem statement, KS and AY performed the experiments, KS and AY analysed the data, KS, PD, and RM wrote the paper.

### Acknowledgment

We acknowledge the funding received by the Science and Engineering Research Board (SB/S2/RJN-105/2017), Department of Science and Technology, India.

### Data Availability

The data that support the findings of this study are available from the corresponding author upon reasonable request.